\documentclass{article}%
\usepackage{amsmath}
\usepackage{amsfonts}
\usepackage{amssymb}
\usepackage{graphicx}

\begin{document}

\begin{center}{\Large \textbf{
The Prediction of Anyons: Its History and Wider Implications\\
}}\end{center}

\begin{center}
Gerald A. Goldin
\end{center}

\begin{center}
Dept. of Mathematics, Dept. of Physics \& Astronomy,\\Rutgers University, New Brunswick NJ, USA

\smallskip
geraldgoldin@dimacs.rutgers.edu
\end{center}

\begin{center}
\today
\end{center}

\section*{Abstract}

Prediction of “anyons,” often attributed exclusively to Wilczek, came first from Leinaas \& Myrheim in 1977, and independently from Goldin, Menikoff, \& Sharp in 1980-81. In 2020, experimentalists successfully created anyonic excitations. This paper discusses why the possibility of quantum particles in two-dimensional space with intermediate exchange statistics eluded physicists for so long after bosons and fermions were understood. The history suggests ideas for the preparation of future researchers. I conclude by addressing failures to attribute scientific achievements accurately. Such practices disproportionately hurt women and minorities in physics, and are harmful to science.

\section{Introduction}
\label{sec:intro}
“Anyons” are quantum particles or excitations, theoretically possible in two space dimensions, with exchange statistics intermediate between bosons and fermions. They are associated with surface phenomena in the presence of magnetic flux. Theoretical applications include explaining the quantum Hall effect, describing quantum vortices in superfluids, and their relevance to quantum computing. In 2020, more than forty years after they were first suggested \cite{LeiMyr1977}, experimentalists succeeded in creating anyonic excitations. The experimental confirmation of their prediction attracted considerable new attention to these fascinating possibilities.

Predicting the anyon required basic changes in our understanding of quantum statistics. The prediction is often incorrectly attributed exclusively to Frank Wilczek, while the first clear predictions were by Leinaas and Myrheim in 1977  \cite{LeiMyr1977} and by Menikoff, Sharp, and myself in 1980-1981 \cite{GolMenSha1980, GolMenSha1981}, from different theoretical perspectives. Wilczek's 1982 work \cite{Wil1982a,Wil1982b} took still a third path to the prediction. He also coined the name ``anyons'' to describe such particles. This article describes the early history of intermediate quantum statistics, including the predecessor ideas that led to those predictions.  Some immediately subsequent insights, often overlooked in citing early published results, are also described \cite{GolSha1983, GolMenSha1983, GolMenSha1985}.

One may ask why the possibility of intermediate statistics took physicists so long to discern, from the time when bosons and fermions were well understood. And why, after fifty years, did three independent predictions occur within such a relatively short time interval? Examining these questions suggests some interesting implications for the teaching of mathematics and physics, relevant to the preparation of future researchers.

Finally, I discuss the systemic failure of scientists’ and journalists’ to attribute scientific achievements accurately. With respect to anyons, omissions of proper citations are current as I write this article, though there is no dispute as to the history. But the “anyon” case is not unique. It provides a case study illustrating a far wider problem. Acknowledgment failure, and its tacit acceptance by the scientific community, does damage far beyond disappointing or hurting a few individuals. Non-recognition disproportionately creates career obstacles for women, Black and other minorities, and scientists in developing countries. Young scientists experience disillusion, even intimidation, and have much to lose by speaking out. And potentially fruitful directions of investigation cannot be pursued when researchers are unaware of their existence.

\section{The idea of the anyon: Why so long?}

\subsection{Anyons and nonabelian anyons}

The idea behind intermediate quantum statistics in two-dimensional space (three-dimensional spacetime) is extraordinarily easy. Let us imagine a pair of indistinguishable particles moving on a two-dimensional surface -- constrained, for instance, to the surface of some material. Suppose they exchange positions, but without actually passing ``through'' each other. They must have done so by moving either clockwise or counterclockwise.  One can characterize any exhange, then, by a winding number: the net number (positive or negative) of counterclockwise windings occurring during the exchange. This feature is specific to two-dimensional space.

If the system is described by a complex-valued wave function $\psi$, one wants $|\psi|^2$ to be invariant under any such exchange. For a single clockwise exchange, $\psi$ might then be multiplied by a complex number of modulus one: $\psi \mapsto [\exp{i \theta}] \psi$. With $\theta = \pi$ we obtain fermions, and with $\theta = 2\pi$ we have bosons. But in two-space, two counterclockwise exchanges in succession are {\it inequivalent} to no exchange. Hence we need not require $\exp{2i \theta} = 1$. For ``any'' value of $\theta$, the predicted outcomes of all physical measurements remain invariant under the exchange.

For three or more indistinguishable particles in two-space, path-dependent exchanges performed in succession no longer commute. Then nonabelian representations of the group describing exchanges, acting by unitary operators on multicomponent wave functions, become possible. Therefore quantum mechanics also allows ``nonabelian anyons''.

With these ideas so simple to describe in an elementary way, why did so many brilliant physicists overlook them for so long? Bosons and fermions were understood in 1924-25; the intermediate statistics of anyons was not explicitly proposed until 1977-82. What led to this idea, so long deferred, becoming one ``whose time had come''?

\subsection{Historical and psychological barriers in physics}

The concept of an {\it epistemological obstacle}, introduced by Bachelard in 1938 \cite{Bac1938} and discussed by Schneider \cite{Schn2014}, is well-known in science and mathematics education research. It refers to prior conceptions that impede the understanding necessary for a breakthrough. As elaborated by Brousseau such obstacles are not a “result of ignorance […] or chance”, but an “effect of prior knowledge that was relevant and had its success, but which now proves to be false, or simply inadequate” \cite{Bro1983}.  The term suggests {\it inevitable} barriers in historical paths of discovery, evolution of conceptual schemes, and ascribing meanings to mathematical representations.

In the psychology of an individual thinker or learner, the parallel notion of a {\it cognitive obstacle} refers to prior knowledge limiting the person's development of new conceptions. As students overcome cognitive obstacles, their stages of learning often recapitulate historical processes. There is an analogy with biology, where the ``ontogeny'' of developing organisms seems to recapitulate the ``phylogeny'' of the species' evolution.

Examples abound of epistemological obstacles in physics and mathematics. To understand that objects without applied forces continue in rectilinear motion; to see space and time as not absolute; to embrace wave-particle duality and the uncertainty principle -- each required physicists to overcome universal categories of experience and abandon previously-successful explanations of observed phenomena. Seeing axioms and propositions as ``self-evident truths'' impeded mathematicians' taking them as arbitrary assumptions characterizing abstract structures. Acceptance of negative numbers, ``imaginary'' numbers, non-Euclidean geometries, and transfinite cardinals, all required overcoming beliefs that these did not ``really exist''.

In physics an empirical source of epistemological obstacle can be the inaccesibility of relevant domains of experiment. Thus frictionless dynamical systems, high velocities approaching light-speed, and observability at the subatomic level, remained difficult or impossible to access for centuries; no driving force from experiment yet demanded conceptual change.

With these ideas in mind, I think it is possible to identify five major epistemological/cognitive obstacles that impeded the prediction of anyon statistics, and all that followed. We can also see how antecedent ideas gradually dismantled those obstacles.

\subsection{Epistemological obstacles to the prediction of anyons}

\smallskip
{\bf{Index vs.~value permutations.}} The first obstacle was the use of index permutations to describe particle exchange. In a configuration of $N$ indistinguishable particles, their positions were labeled with subscripts (indices) $1,...,N$; with the wave function written $\psi(x_1, ..., x_N)$. A permutation $\sigma_{(ij)}$ exchanged ``particle $i$'' with ``particle $j$'' -- an index permutation. With this meaning of exchange, there is conceptually no physical path of exchange -- the particles are relabeled abstractly. Writing $|\psi(x_1, ..., x_N)|^2$ = $|\psi(x_{\sigma(1)}, ..., x_{\sigma(N)})|^2$ asserts invariance under index exchange -- reintroducing indistinguishability after labeling the particles distinguishably.

Value permutations, in contrast, make reference to the coordinates locating the particles being exchanged; i.e., their positions in the physical space. The permutation $\sigma_{(ij)}$ exchanges the particle in the $i$-th location with the one in the $j$-th location. This description requires an ordering of points in physical space. The arbitrariness of that (which is actually no stronger objection than the arbitrariness of indexing) is one reason for conceptual difficulty in letting permutations act on coordinate values rather than indices. Nevertheless, exchanging actual particle positions allows one to focus on possible paths of exchange.

Note that introducing labels or adopting a coordinate system is {\it always} arbitrary. By itself, it leaves the physics invariant. A symmetry property of coordinates is not a physical symmetry, and in principle provides no new physical information. Physical insight comes when the symmetry group is understood to act on the system itself, with some system properties identified as invariant under the symmetry. This distinction, while essential and obvious, is often easily overlooked in the language we use to discuss group theoretical applications in physics.

\smallskip
\noindent
{\bf{Coordinate space vs.~configuration space.}} A second obstacle inheres in considering ordered $N$-tuples at all. A configuration of indistinguishable particles is actually an {\it unordered} $N$-point subset of the physical space. To see statistics as arising from paths of exchange requires a focus on configuration space topology. But an unordered set allows no description of particle exchange -- hence the introduction of indices and imposition of invariance under exchange, obscuring that focus. A different idea was needed. Note that the expression $\Delta_N^{(d)} = [(\mathbb{R}^d)^{\times N}-D]\,\mathit{mod}\,S_N$ for $N$-particle configuration space in $\mathbb{R}^d$, where $D$ is the ``diagonal'' consisting of $N$-tuples for which $x_i = x_j$ for some $i \not= j$, and $S_N$ is the symmetric group, seems to reflect the historical conception. It is much simpler to write $\Delta_N^{(d)} = \{\gamma \subset \mathbb{R}^d| \,\mathrm{card}(\gamma) = N\}$.

\smallskip
\noindent
{\bf{Continuous single-valued wave functions.}} A third obstacle was the assumption over many years that wave functions on coordinate space must be continuous and single-valued. This perhaps reaches as far back as the Bohr atom, where electrons were posited to occupy fixed circular ``orbits''. A continuous wave describing this would have an appropriate period, forbidding self-interference and leading naturally to quantization. In subsequent models based on the Schr{\"o}dinger equation, single-valuedness and continuity provide a natural framework for quantization of energy and momentum. The conception of $\psi$ as a kind of physical field modeled on points in physical space (albeit configurations of such points) also seemed to demand single-valuedness. Thus if $\psi \mapsto [\exp{i \theta}] \psi$ under a single exchange, of necessity we need $\exp{2i \theta} = 1$. The ``simple'' idea provided above is then easily dismissed as fallacious.

\smallskip
\noindent
{\bf{Established empirical knowledge: fermions and bosons.}} The dramatic achievements of quantum theory with just two types of particles posed a fourth obstacle. Successes included the Pauli exclusion principle for fermions, standing behind the periodic table of elements and explaining in principle chemical reactions. They included the phenomenon of Bose-Einstein condensation; also local quantum field theories where bosons are the quanta carrying fundamental forces of nature. No experiments compelled inquiry into more exotic possibilities.

\smallskip
\noindent
{\bf{Axiomatic quantum theory.}} Finally, a fifth obstacle inhered in the understanding, achieved through axiomatic relativistic quantum field theory, of how some fundamental physical laws follow from basic assumptions. The Wightman axioms \cite{StrWig1964} included the proposition that space-like separated fields either commute or anticommute. These axioms led to rigorous proofs of $PCT$-invariance and the spin-statistics connection. Of course, the axioms encoded widely-shared beliefs about the properties quantum fields should have. But working from fixed axioms does create an intellectual context where they are no longer questioned; only their implications are explored. Bose and Fermi statistics exclusively were thus firmly embedded in the foundations of physics.

\section{Ideas antecedent to the prediction of anyons}

Important antecedent ideas, developed over several decades, eventually overcame these obstacles to intermediate statistics.

\smallskip
\noindent
{\bf{Intermediate occupation number statistics.}} As early as 1940-1942, Gentile \cite{Gen1940, Gen1942} explored the possibility of hypothetical occupation number statistics other than those of fermions or bosons, where  an intermediate, finite number of particles could be permitted to occupy the same quantum state. He drew some possible consequences for the theory of superfluidity.

\smallskip
\noindent
{\bf{Topology in quantum mechanics.}} In 1959 Aharonov and Bohm \cite{AhaBoh1959} considered charged particles excluded by an infinite potential barrier from a cylindrical region, behind which a current in a tightly-wound infinitely-long solenoid sustains a magnetic flux. Quantum theory then predicts shifts in the energy and kinetic angular momentum spectrum of the particles, despite the absence of a physical magnetic field in the accessible region. Their paper, conceived as a {\it Gedanken} experiment, evoked much controversy as to its meaning. Though known as the ``Aharonov-Bohm effect'', a similar but much less noticed proposal had actually been published in 1949 by Ehrenberg and Siday \cite{EhrSid1949}, as Hiley describes in a historical article \cite{Hil2013}.  The earlier work is highlighted in the Wikipedia entry on the subject \cite{Wiki2022a}, which is how I learned of it just last year. The (Ehhrenberg-Siday)-Aharonov-Bohm effect pointed to the role of the topology of the space in which quantum particles move -- particularly, but not exclusively, when charged particles circle excluded regions of magnetic flux. The possibility of multivalued wave functions needed to be entertained -- though challenged, it began to achieve legitimacy.

Following on these ideas, in the context of the Feynman path-integral fomulation, physicists considered homotopy classes of trajectories from initial to final particle configurations. Here one moves from the topology of physical space to that of configuration space. Schulman \cite{Schu1968} in 1968 proposed a model for the topological origin of particle spin. In 1971, Laidlaw and (C\'{e}cile) DeWitt \cite{LaiDeW1971} deduced the topological origin of Fermi and Bose exchange statistics. In a footnote to their result, they remarked that in two space dimensions there seemed to be additional possibilites for quantum statistics, but they did not pursue this. In 1972, Dowker \cite{Dow1972} provided a more general discussion of quantum theory on multiply-connected spaces.

\smallskip
\noindent
{\bf{Parastatistics, kinks, etc.}} During the 1950s and 1960s, other paths led to some generalizations of exchange statistics. In 1953, Green \cite{Gre1953} obtained parastatistics from trilinear brackets of quantum fields (combining canonical commutation and anticommutation relations). This work, with resulting investigations of symmetrization in 1964-1965 by Messiah and Greenberg \cite{MesGre1964} and Girardeau \cite{Gir1965}, brought in higher-dimensional, nonabelian representations of $S_N$. As concrete alternatives to Bose and Fermi statistics, parastatistics evoked unfulfilled conjectures that fundamental particles such as quarks might satisfy them. In 1968 Finkelstein and Rubinstein \cite{FinRub1968} suggested more general possibilities for the spin/statistics relation for ''kinks'' in the context of quantized nonlinear fields, by admitting double-valued state functionals.

\smallskip
\noindent
{\bf{Braid groups and Yang-Baxter relations.}} As different models in quantum field theory were invented and studied, braid groups began to enter into consideration in the late 1960s and 1970s. In two-dimesional models with soliton fields they found a place through the Yang-Baxter equation in articles by Streater and Wilde \cite{StrWil1970} and Fr{\"o}hlich \cite{Fro1976a,Fro1976b}  Related work was published by Klaiber \cite{Kla1968}, Souriau \cite{Sou1970}, Kadanoff and Ceva \cite{KadCev1971}, and Wegner \cite{Weg1971}.

\smallskip
\noindent
{\bf{Group representations and current algebras.}} In parallel with these developments, unitary group representations came into their own as pillars of quantum theory. Wigner, Mackey, and numerous others established their fundamental role \cite{Wig1939,Var2008}.  In strong interaction physics and the theory of fundamental particles, dramatic findings confirmed the predictive power of $SU(3)$ \cite{GelNee1968}.  This led eventually to unification of the electroweak and strong forces via $SU(2) \times U(1) \times SU(3)$ gauge theory in the ``Standard Model''.

Moving from Lie groups to Lie algebras, Adler and Dashen \cite{AdlDas1968} made a strong case in the 1960s for current algebras in fundamental particle physics. In 1968 Dashen and Sharp \cite{DasSha1968} proposed a certain highly singular, local current algebra describing nonrelativistic quantum field theory. Their goal was to describe hadrons by gauge-invariant quantities such as densities and currents, rather than gauge-dependent quantum fields. Behind this work stood earlier ideas of Haag and Kastler \cite{HaaKas1964} and others on algebras of local observables.

In the late 1960s, I was able to regularize and exponentiate Dashen and Sharp's algebra to obtain an infinite-dimensional group, and establish a framework for studying its unitary representations \cite{Gol1971}. The group is the natural semidirect product of a diffeomorphism group {\it Diff}$_0(\mathbb{R}^3)$ of the physical space (describing flows generated by momentum density operators), with an additive group of scalar functions on $M$ (describing exponentiated mass density operators). Then in collaboration with Grodnik, Powers, and Menikoff \cite{GolSha1970,GolGroPowSha1974,Men1974,MenSha1975,MenSha1977}, Sharp and I found applications confirming the fundamental role of this group and its Lie algebra. Across the 1970s, Menikoff, Sharp, and I successfully extended Mackey's method of induced representations of locally compact Lie groups, to obtain a class of unitary representations of diffeomorphism groups \cite{GolMenSha1980}.  It followed from our work that Bose and Fermi exchange statistics could be understood as inequivalent unitary representations of our semidirect group, induced by representations of $S_N$. And $S_N$ entered naturally as the fundamental group of $N$-particle configuration space. This provided a new, wholly kinematical perspective on the earlier work of Laidlaw and DeWitt \cite{LaiDeW1971}, which had been based on Feynmann paths; and it led to intermediate quantum statistics in two-space.

The roles of $SO(3)$ and $SU(2)$ in describing orbital and spin angular momentum were of course long known at the time of all this work. A 1976 paper by Martin \cite{Mar1976} proposed a model for the Aharonov-Bohm effect based on rotation generators in two-dimensinal space. Here she presaged group-theoretically the fractional statistics subsequently predicted for anyons.

\section{Intermediate statistics: Three independent predictions}

By the mid- to late 1970s requisite ideas were in place, the obstacles mostly removed.

In 1977, Leinaas and Myrheim \cite{LeiMyr1977} presented the first clear prediction of quantum exchange statistics interpolating Bose and Fermi statistics in two space dimensions. They based their analysis on  Schr{\"o}dinger quantization of particle dynamics, using the topology of Feynman paths. They drew a connection with electromagnetism, noting the singularity in configuration-space associated with the coincidence points of particles. In 1978 Leinaas \cite{Lei1978} suggested a model based on charged-particle/monopole composites.

This picture did leave open the issue of whether Feynman paths might ``cross'' -- i.e., can two particles ``pass through'' each other as the quantum configuration evolves? One might then need a hard-core, singular repulsive potential for intermediate statistics. In the Aharonov-Bohm setup, the nontrivial topology was established by introducing an infinite barrier to exclude the charged particles from a region of space; would such exclusion be necessary here?

Menikoff, Sharp, and I published our prediction of intermediate statistics in 1980-81, not yet aware of the Leinaas-Myrheim papers. Our findings confirmed theirs but assumed less, being kinematical rather than dynamical \cite{GolMenSha1980,GolMenSha1981}.  Studying the Aharonov-Bohm setup with our local current algebra meant representing the group of diffeomorphisms of a non-simply connected space. We had already established a foundational role for unitary representations of {\it Diff}$_0(\mathbb{R}^3)$ in classifying quantum systems; now we found -- to our surprise at the time -- that the unitary representations of {\it Diff}$_0(\mathbb{R}^2)$ included intermediate exchange statistics. Thus we did not obtain our results by quantizing a classical system, but by rigorously pursuing fundamental group-theoretic methods. Discovering intermediate statistics culminated 15 years of research.

A number of things became clear from our work. One obtains directly the shifted spectrum of self-adjoint kinetic angular momentum operators associated with anyons. Wave functions are single-valued on the true configuration space for indistinguishable particles; the exchange statistics is established by the operators describing local observables. Coincidence points are excluded necessarily and not arbitrarily; no repulsive potential is necessary. The (illusory) multi-valuedness of wave functions reflects an equivalent representation on the Hllbert space of {\it equivariant} wave functions on the universal covering space of configuration space. Equivariance is with respect to its fundamental group (first homotopy group). The inner product is defined by integration on configuration space, not on the covering space -- an essential idea, because there are in general infinitely many sheets to the covering space. 

In 1982 Wilczek, who by his second article that year knew of the earlier articles, published his own, independent prediction \cite{Wil1982a,Wil1982b}. His systematic investigation of fractional quantum numbers suggested fractional-spin particles in two dimensions. He modeled this with charged particles bound to units of magnetic flux orthogonal to the surface confining the particles -- like miniature ``Aharonov-Bohm solenoids'' with net charge. Wave functions describing such particle/flux tube composites pick up the intermediate phase $\exp{i\theta}$ in a single counterclockwise exchange. His name ``anyons'' expresses that $\theta$ can take ``any'' value between $0$ and $2\pi$.

Wilczek recognized and advocated for anyons' theoretical importance, especially as applications were found to understanding the quantum Hall effect. Twenty-three of the most impactful articles from 1983 to 1990, including seven by Wilczek and his collaborators, are reprinted in his 1990 book, {\it Fractional Statistics and Anyon Superconductivity} \cite{Wil1990}.  Space here does not permit my citing them; the reader is referred to \cite{Wil1990}. This influential volume provided a valuable resource for researchers and reviewers of the field across the next decades. 

Some immediate, fundamental consequences of my work with Menikoff and Sharp at Los Alamos were also published during the 1980s. In 1983, we presented a rigorous kinematical framework for the fractional spin of anyons, including the first (as far as I can determine) explicit identification of the braid group $B_N$ as the homotopy group whose unitary representations govern $N$-anyon exchange statistics \cite{GolSha1983,GolMenSha1983}.  In 1985 we first predicted nonabelian anyons \cite{GolMenSha1985}, described by wave functions equivariant under higher-dimensional unitary representations of $B_N$. This parallels the earlier idea of parastatistics for particles in $\mathbb{R}^3$, where the homotopy group is $S_N$. We also pointed out in 1985 that systems of distinguishable  particles in two-space are described by wave functions equivariant for the group of colored braids. These wave functions can pick up intermediate phases as particles fully circle each other, without exchange. Our conclusion about nonabelian anyons was contrary to the expectation expressed by Wu in 1984 \cite{Wu1984} that a ``general theory'' would include only one-dimensional representations of the braid group; we published it as a response to Wu's paper.

Many further applications of anyon theory followed across the decades; not only in physics (e.g., to the quantum Hall effect and to quantum vortices), but also in the burgeoning field of quantum computing. Two years ago, more than four decades after their first prediction, two groups of experimental physicists announced success in creating and observing anyonic excitations \cite{Bar2020,Nak2020}, stimulating wide interest.

\section{Implications for the education of future physicists}

This retrospective on the first predictions of anyons, the historical obstacles that delayed them and the antecedent research that removed those obstacles, suggests some more general considerations for future (or current) physicists. I would like to offer a few thoughts on the topic of education, before turning to the issue of citation integrity in physics. Epistemological and cognitive obstacles are most likely present now, though we may be unaware of them. In teaching university-level theory -- classical mechanics, electricity and magnetism, optics, quantum mechanics, relativity, thermodynamics -- one goal should be to facilitate scientifically sound exploration that can penetrate or even overturn prevailing conceptions.

How can we foster students' questioning of established constructs and encourage generation of new ones? What best enables a student to identify tacit assumptions and make them explicit? What tools  help sudents overcome personal cognitive obstacles? I think that answers consist, in part, in exploring what it means for a student (or researcher) to ``really understand'' a newly-studied concept in physics or mathematics.

Consider the skills normally emphasized (or not) in teaching theory to physics students:

\smallskip
\noindent
{\bf{Textbook problems.}} Solving progressively more complex problems based in established theory, using standard techniques, is central to most physics courses. Students develop their understanding of theoretical constructs by applying them to situations where they fit directly.

\smallskip
\noindent
{\bf{Nonroutine problems.}} Many physics courses incorporate some problems that are less routine. Valuable heuristic methods and general strategies typically apply in such contexts. The best students gradually acquire them, though they may or may not be discussed explicitly. Examples include examining special cases, testing limiting cases or idealized cases, creating multiple representations, establishing and using insightful notation, choosing a helpful coordinate system, finding hidden symmetry, exploiting units of measurement, carefully distinguishing what is happening physically from its mathematical description, and so forth. Most such strategies pertain to mathematics as well as to physics. 

But sometimes we bypass the thinking process: presenting students with notation, suggesting the desired representation, providing the coordinate system, and pointing the way to insight rather than allowing the experience of discovery. If we consistently dismantle obstacles, we may limit students' development of powerful methods for breaking through them.

\smallskip
\noindent
{\bf{History of discoveries.}} Physics teaching typically includes stories about discoveries and breakthroughs, how historic experiments forced reconsideration of previously accepted theories. I think there is much more we can do.  We can explore the  {\it thinking process} that led to a new theory. We can identify critical epistemological/cognitive obstacles delaying its invention. We can ask what philosophical or metaphysical assumptions may have impeded the idea, and how one might have noticed these earlier. We can explore alternative theories that did not pan out.

\smallskip
\noindent
{\bf{Students as inventive theorists.}} In introducing the foundational ideas behind a new concept, we might seek to engage students in thinking as original theorists. Before presenting established theory, students can offer their own conjectures and explore their consequences. The goal is not to see who is ``right" and who is ``wrong'', but to foster students' creative theorizing. We can study rival theories and abandoned theories, to consider how one evaluates a scientific idea as valid or worth pursuing.  I favor posing the following question to students and to ourselves: ``If no one had ever seen this idea before, or if you had never previously encountered it, can you imagine how you personally might have invented it?'' 

\section{A sequel to discovery: Citation omission and its consequences}

Scientific research is an intimate activity. Disappointments and frustrations are interspersed with occasions of insight and satisfaction. One rarely achieves in full what one aspires to, but strives to fulfill the ideals of one's teachers and mentors. One values their words of encouragement, their belief in one's ability to contribute to understanding the natural world. One develops friendships and shared memories of collaborative success. One does what one loves.

A young mathematical physicist might realistically hope to create some interesting new models, to place some well-established physics on a more rigorous foundation, to unify previously disparate phenomena, or to show how some physical effects result from more fundamental laws. But perhaps the highest aspiration of the young theorist might be to predict a wholly new, unsuspected phenomenon -- and to see that prediction confirmed.

Four decades after their prediction, anyonic excitations were observed by two experimental groups \cite{Bar2020, Nak2020}.  This should be a deep source of satisfaction to all of us who put forth the prediction. Developing the theory that led to my own group's prediction took $15$ years, during difficult times professionally. But both groups of experimentalists, apparently unaware of our early published work, cited only the articles by others. Shortly thereafter, widely circulated science magazines repeatedly attributed the prediction of anyons exclusively to Wilczek \cite{Orn2020,Naj2020,Cot2021,Dre2021}, who had shared the Nobel Prize in 2004 for his earlier work on quark confinement and asymptotic freedom. And the related fundamental insights about anyons that we were first to publish remain wholly unacknowledged.

How this came about, and how it continues today, are also part of the history of anyons. We have learned that the problem is systemic, stemming from omissions inadvertent and intentional, and magnified in impact by open journalistic practice contrary to published policies on fairness in reporting. But citation inequity does more than disappoint a few individual researchers. It has far wider implications for science.

It should not be necessary for me to emphasize that in describing this history, there is no intent whatsoever to diminish the fact or importance of Wilczek's substantial contributions to the physics of anyons. I am aware of no dispute among any of us regarding the authorship, the priority, the content, or the originality of the published results. 

\subsection{Anyons: A case study in systematic citation omission}

Between 1982 and 1989, omission of proper reference to our articles was widespread. Most of these omissions were inadvertent, as researchers relied in their bibliographic research on the citations in earlier papers. But some, beginning with the most often-cited article \cite{Wil1982b}, were intentional and consistently maintained. The effects were immediate and enduring. Among all the numerous citations in the $23$ articles from 1983 to 1990 reprinted in Wilczek's influential 1990 book \cite{Wil1990}, there is just one citation of an article of ours. In the period that followed, this greatly limited awareness of our work by those for whom the book served as a major resource.

Acknowledgment as it relates to one's own research is a highly personal and difficult subject to address factually and objectively. It is a remarkable experience to be written out of history before one's eyes, year after year. Our work appeared in leading refereed journals. Our approach to intermediate statistics was novel -- wholly group-theoretical in its foundation -- and independently developed. Some findings that became widely known we were first to publish. As the omissions persisted, my colleagues and I made every effort to acquaint others with our published articles. At the time, internet and email did not exist. Correspondence was through letters -- slow, difficult to write, always polite, and frequently unanswered. It made no difference.

In retrospect, our efforts in the 1980s were not sufficiently aggresssive. We relied on the good will and integrity of the research community. We sought to be scrupulous in acknowledging the work of others, and expected proper reference to our findings. But we saw how fame, connection, and power relationships influence acknowledgment choices. The absence of information availability closed off interest in some research directions. Theoretical methods pursued were those people knew of -- which sometimes led to insights we had already obtained by other means.

Then in 1989, \emph{Physics Today} published an extensive article about anyons \cite{Khu1989} attributing their discovery to Wilczek, mentioning Leinaas and Myrheim's article briefly, and omitting all reference to our work. My colleagues and I requested a correction. Somewhat to our surprise, there ensued an extraordinary battle -- at heavy personal cost -- before we achieved publication of the proper attributions. Eventually, the error was corrected. A detailed letter, authored by Biedenharn, Lieb, Simon, and Wilczek \cite{Bie1990}, described our contributions accurately and succinctly. Wilczek included a sentence in his book \cite{Wil1990} (p. 105) citing our three earliest papers. He invited our contribution to a special issue of {\it International Journal of Modern Physics B} he was then editing \cite{GolSha1991a}. There had never been an actual dispute regarding the sequence of discoveries, and it seemed that after eight years the acknowledgment issue was resolved.

But in 1991, an article about anyons in {\it Scientific American}, authored by Wilczek, credited him with the discovery without reference to any prior predictions \cite{Wil1991}  This was corrected by letter from Sharp and me, after some struggle \cite{GolSha1991b}.

Over the next thirty years, citations appeared only sporadically. During this period, I came to believe that had the internet existed back in the 1980s, such consistent failures of attribution could never have occurred. Since 2020, Sharp and I have learned how untrue that is.

With the recent experimental findings, fame seems to have triumphed decisively over journalistic accuracy in science reporting. Omissions, no longer inadvertent, are openly deliberate. Efforts at correction are ignored or refused. In two successive articles, {\it Discover Magazine} credited only Wilczek with the prediction \cite{Orn2020,Cot2021}  They disregarded detailed communications, refusing to post a simple update to their featured on-line articles \cite{GolSha2021a,GolSha2021b}  In this, the editors chose to violate explicit canons of ethics in journalism. For example, the policy of the {\it Washington Post} states:
\begin{quote}
“Fairness results from a few simple practices: No story is fair if it omits facts of major
importance or significance. Fairness includes completeness.
...
No story is fair if it consciously or unconsciously misleads or even deceives the reader. Fairness includes honesty — leveling with the reader.” \cite{WashPost}
\end{quote}
{\it Quanta Magazine} also published two features, crediting only Wilczek with the prediction of intermediate statistics \cite{Naj2020, Dre2021}  Its editors did not respond to repeated efforts to communicate our request for a correction. Thus in major journalistic outlets, it is as if neither Leinaas and Myrheim nor our group had participated at all in the research.

Researchers identifying appropriate citations in a specialized domain of physics often consult the relevant Wikipedia entry, and follow up with academic sources. I do so myself. Though not always 100\% accurate or complete in describing the physics, Wikipedia is important.

At Wikipedia, the ``anyon'' entry \cite{Wiki} (discussing anyons, nonabelian anyons, topology, the braid group, and fractional spin) omits all indication of my colleagues' and my correct predictions about these topics. An anonymous editor (screen name ``HouseofChange'') has expunged every correct citation entered by at least two independent experts. This editor claims that the absence of earlier citations proves our work to be irrelevant, and characterizes the {\it Journal of Mathematical Physics} as ``obscure''. It is the opposite stance to Wikipedia's informative entry about the ``Aharonov-Bohm effect'' mentioned above, acquainting readers with the prior work of Ehrenberg and Siday. When removing the citations by other editors, ``HouseofChange'' leveled false accusations against them, and asserted untruths about Sharp and myself. Most or all exchanges are on Wikipedia's ``talk'' feature. One unknown person still ensures that those interested in the physics of anyons should never know of our prediction, or of the group-theoretical and current-algebraic methods that led to it.

To sum up, in the nearly two years since the experimental confirmation of anyonic statistics, the most strenuous efforts possible have produced not one additional correction or footnote in the information generally available to the public. 

\subsection{Citation inequity in physics: Untruth and its consequences}

{\it Why does this matter?} If one is not directly affected, one might be a little amused by the importance accorded here to brief mention and a few footnotes, and the fierce resistance to it. But there is a wider perspective. Denial of information slows research and its attendant benefits -- an intangible loss to the entire scientific community. But this is far from the only damage. Acknowledgment failure profoundly undermines fairness and equity.

Egregious examples of overlooking women in physics for top honors are well-known: Chien-Shiung Wu's pioneering work on parity conservation violation, disregarded in the 1957 Nobel Prize award to Lee and Yang; Jocelyn Bell's crucial role in discovering pulsars, unrecognized in the 1974 Nobel to Hewish and Ryle. But for every such inequity at the highest levels, countless examples occur in less dramatic contexts. Substantial citation inequity toward women in phyiscs on a wide scale has been recently documented \cite{Tei2022}, while our field remains overwhelimingly male.

Unfairness toward women is not the only form of discrimination citation inequity takes. It hurts Black and other minority scientists, those in developing countries, and early career researchers who have much to lose by speaking out -- all who lack influential connections. In his 2021 book {\it Fear of a Black Universe} \cite{Ale2021}, physicist Stephon Alexander highlights the experiences of Black physicists who are marginalized. Interviewing with the {\it The Guardian} \cite{Gua}, he notes how citation inequity affected his mentor, physicist Jim Gates, whose contribution to supersymmetry together with Nishino was disregarded for at least ten years: 
\begin{quote}
``I was there when Jim realized that the work was not cited and he wrote one of the authors directly. Then they cited it, but it was kind of too late. That’s why I wrote about it in this book, to celebrate that it was [Gates and Nishino’s] discovery. ... That is exactly the phenomenon that Black people experience in other fields where we’re not supposed to occupy these spaces.''
\end{quote}

Non-recognition of scientific achievement and the resulting career obstacles may be affecting younger readers of this article, even as I write. Success in pursuing a long-term, original project is never assured. Less-traveled paths may mean more limited opportunites. To predict a new, unsuspected phenomenon in one's early career risks evoking skepticism, particularly if the time for the prediction is not yet ripe. When the best results go unrecognized and unacknowledged, consequences can be serious. Disillusionment, alienation, and discouragement about risk-taking set in, and may call young physicists' love of science into question. And citations matter greatly in university tenure and promotion decisions. 

Most important of all is the issue of our community's commitment to scientific integrity. Does the physics community (or more generally, the wider scientific community) recognize excellent research through quiet hard work, or do we value prominence, fame, or promotional ability more? Honest and thorough acknowledgment of prior research should be a standard for every scientific publication, taken as seriously as we now take other standards of integrity such as truthful representation of data and authenticity of authorship. Genuinely inadvertent omissions, while probably inevitable, are easily corrected with on-line updates. When untruth is deemed to be minor and correcting it unimportant, or when intentionally incomplete attribution goes unchallenged, the slope is slippery. In science reporting, purposeful unfairness must be denounced. Tolerance for it endangers the very value we place as scientists on truthfulness. 

\subsection{A personal note}

Sharp and I have been extremely privileged. Encouraged to pursue science, we received superb educations and graduate fellowships at elite U.S. universities. We had successful careers in research. We received honors and awards in physics. Research risks taken led to challenging obstacles in the 1970s, but we were able to see these through and continue as active scientists.

In retrospect it is distressingly clear that only our status and connections in 1989-91 allowed us to achieve the corrections published in {\it Physics Today} and {\it Scientific American}  Less connected researchers would have had no chance. And today the obstacles to correcting unfair omissions seem vastly greater! With all our credentials and refereed publications, we cannot achieve passing mention by science journalists drawn to a more famous person, or a few footnotes in Wikipedia for undisputedly original findings. What can others with fewer resources do?

Thus Sharp and I came to feel it is our obligation to speak out, not just for ourselves but for science. The outcome will not benefit us materially, nor will continued disregard harm us furher. A public stand is necessary to address a systemic problem of integrity in physics.

\section{Conclusion}

We have reviewed the history of ideas leading to the prediction of anyons. Epistemological obstacles stood in the way; research across decades helped break through them, leading to three independent predictions. Close study suggests the value of educating future theorists in the history of discoveries, encouraging them to question assumptions and invent alternate explanations. We have also described systemic citation inequity, and ongoing dishonest journalism. Such practices have adverse consequences not only for individuals but for science.

My faith is that the physics community fundamentally values truth and integrity in scholarship, and that inequitable practices contrary to these values will become unacceptable. Hopefully this article contributes meaningfully toward that goal.

\section*{Acknowledgments}

Jürg Fröhlich (ETH Zürich) and Douglas Lundholm (Uppsala University) suggested valuable references antecedent to the prediction of anyons. I am grateful to the organizers of the Group34 Colloquium for the opportunity to speak in the 50th Anniversary Special Session.  Finally, I would like to acknowledge David H. Sharp for his seminal contributions to the research reviewed here, and for our life-long collaboration.

\end{document}